\documentclass{eptcs}
\usepackage{amsfonts}
\usepackage{amssymb}
\usepackage{amsmath}
\usepackage{prooftree}
\usepackage{array}
\usepackage{breakurl} 
\usepackage{mdwlist}

\newcommand{\RR}{\mathcal{R}}

\newcommand{\EE}{\mathcal{E}}
\newcommand{\G}{\Gamma}

\newcommand{\SALTA}[1]{}

\newcommand{\XX}{\mathcal{X}}
\newcommand{\TV}{\mathcal{TV}}
\newcommand{\SV}{\mathcal{SV}}

\newcommand{\xx}{\widetilde{x}}

\newcommand{\Ltrans}[1]{\Longrightarrow}
\newcommand{\Loop}[1]{\left(#1\right)^L}

\newcommand{\into}{\ensuremath{\,\rfloor}\,}
\newcommand{\pipe}{\ensuremath{\,|\,}}
\newcommand{\phole}{\square}

\newcommand{\pair}[2]{(#1,#2)}
\newcommand{\dom}[1]{dom(#1)}

\newcommand{\agr}{\quad\big|\quad}

\newcommand{\eq}{::=}
\newcommand{\CL}{CL}
\newcommand{\E}{E}
\newcommand{\M}{M}
\newcommand{\I}{I}
\newcommand{\R}{R}
\newcommand{\labelSet}{L}
\newcommand{\act}{\genMethod{action}}
\newcommand{\ass}{\genMethod{ass}}%ociation}}
\newcommand{\dis}{\genMethod{dis}}%sociation}}
\newcommand{\inm}{\genMethod{in}}
\newcommand{\outm}{\genMethod{out}}
\newcommand{\obj}{\texttt{Object}}
\newcommand{\mol}{\texttt{Molecule}}
\newcommand{\sug}{\texttt{Sugar}}
\newcommand{\enz}{\texttt{Enzyme}}
\newcommand{\hyd}{\texttt{Hydrolase}}
\newcommand{\glhyd}{\texttt{GlycosideHydrolase}}
\newcommand{\enzcomp}{\texttt{EnzymeComplex}}
\newcommand{\bioobj}{\texttt{BioObject}}
\newcommand{\por}{\texttt{Porin}}
\newcommand{\lab}{\texttt{Label}}
\newcommand{\phoiso}{\texttt{ph-iso}}
\newcommand{\glu}{\texttt{glu-6-ph}}
\newcommand{\fru}{\texttt{fru-6-ph}}
\newcommand{\this}{\texttt{this}}
\newcommand{\ok}{\  \texttt{OK}}
\newcommand{\okIn}[1]{\  \texttt{OK in }#1}
\newcommand{\tol}[1]{\  \overset{#1}\to\  }
\newcommand{\ruleName}[1]{\textrm{(#1)}}

\newcommand{\genMethod}[1]{#1}
\newcommand{\genClass}[1]{#1}

\newcommand{\CTt}{CT}
\newcommand{\CT}[1]{\CTt(#1)}

\newcommand{\multiple}[1]{\overline{#1}}
\newcommand{\tolr}[2]{\  \overset{#1}{\underset{#2}\rightleftharpoons}\  }
\newcommand{\mtype}[2]{mtype(#1,#2)}
\newcommand{\mbody}[2]{mbody(#1,#2)}
\newcommand{\subtype}[2]{#1<:#2}
\newcommand{\typeAss}[2]{#1:#2}
\newcommand{\methoddef}[2]{#1(#2)}
\newcommand{\clssdef}[2]{\texttt{class } #1 \texttt{ extends } #2}
\newcommand{\clss}[3]{\texttt{class } #1 \texttt{ extends } #2 \texttt{\{}#3\texttt{\}}}
\newcommand{\genVariable}[1]{#1}
\newcommand{\method}[3]{#1(#2) \  #3}
\newcommand{\invoc}[3]{#1.#2(#3)}

\title{A Minimal OO Calculus for Modelling Biological Systems\thanks{This work was partly funded by the project BioBIT of the Regione Piemonte.}}
\author{Livio Bioglio
\institute{Dipartimento di Informatica\\
 Universit\`a di Torino\\
  Torino, Italy}
\email{biogliol@di.unito.it}
}
\def\titlerunning{A Minimal OO Calculus for Modelling Biological Systems}
\def\authorrunning{L. Bioglio}

\begin{document}
\maketitle
\begin{abstract}
In this paper we present a minimal object oriented core calculus for modelling the biological notion of type that arises from biological ontologies in formalisms based on term rewriting. This calculus implements encapsulation, method invocation, subtyping and a simple form of overriding inheritance, and it is applicable to models designed in the most popular term-rewriting formalisms. The classes implemented in a formalism can be used in several models, like programming libraries.
\end{abstract}

\section{Introduction}
In biology, homogeneous biological entities are usually grouped according to their behaviour. Enzymes are proteins that catalyse (i.e. increase the rates of) chemical reactions, receptors are proteins embedded in a membrane to which one or more specific kinds of signalling molecules may attach producing a biological response, hydrolases are enzymes that catalyse the hydrolysis of a chemical bond, and so on. Such classification is greatly behaviour-driven: the lactase is a hydrolase, then its peculiarity with respect to the other biological entities is that it catalyses the hydrolysis of a particular molecule. It suggests Computer Science types: every biological entity can be classified with a type, containing the sound operations for it. These operations describe only the general behaviours, that may be modelled by means of different formalisms. Like Computer Science types, these Biological types are used to check the correctness of the chemical reactions. In fact, lactase is not just a hydrolase, but a glycoside hydrolase, i.e. it catalyses the hydrolysis of the glycosidic linkage of a sugar to release smaller sugars. If in the system the substrate or the products are not sugars, somewhere there is an error. Moreover, in Biological types we can recognize a subtype relation. Lactase hydrolyse the lactose, that is a disaccharide. Since the disaccharide is identified as a subtype of sugar, the hydrolysis operation associated to the glycoside hydrolase type is correct.\\
Many formalisms originally developed by computer scientists to model systems of interacting components have been applied to Biology: among these, there are Petri Nets~\cite{MatDoiNagMiy00}, Hybrid Systems~\cite{AluBelKumMin01}, and the $\pi$-calculus~\cite{Cur_etal04,RegSilSha01}. Moreover, new formalisms have been defined for describing biomolecular and membrane interactions, for example~\cite{BarMagMilTro06a,Car05,CCD04,DanLan04,PriQua05,Reg_etal04,Pau02}. Even if types are used by biologists and studied by computer scientists, curiously they are usually not implemented in Computer Science biological models. Despite the number of formalisms developed by computer scientists and applied to model biological systems, just in the last few years there has been a growing interest on the use of type disciplines to enforce biological properties. In~\cite{FS08} three type systems are defined for the Biochemical Abstract Machine, BIOCHAM (see~\cite{BioCHAM}). In~\cite{DGT09a} a type system for the Calculus of Looping Sequences (see~\cite{BarMagMilTro06a}) has been defined to guarantee the soundness of reduction rules with respect to the requirement of certain elements, and the repellency of other ones. Finally, in~\cite{CT10}, group types are used to regulate compartment crossing in the BioAmbients framework~\cite{Reg_etal04}. However, none of them exploits the similarities between the types in Biology and in Computer Science.\\
In this paper we present a minimal object oriented core calculus for term-rewriting formalisms, i.e. formalisms based on term rewriting, that models the notion of types used in biology as above described. We implement only the object oriented paradigm skills that, in our view, are basic in modelling biological systems, that is encapsulation, method invocation, subtyping and a simple inheritance. The purpose of this calculus is to facilitate the organizations of rules, and to improve their re-use in the model, or even in other models. By means of subtyping, for example, modellers create a class hierarchy, that can be used in different models like programming libraries: classes and methods are created by expert researchers, but they can also be used by raw users.\\
The remainder of this paper is organized as follows. In Section \ref{sec_def} we formally present the core calculus. In Section \ref{sec_exa} we propose classes explaining some enzyme behaviours. In Section \ref{sec_use} we apply our framework to two term-rewriting formalisms, the Calculus of Looping Sequences~\cite{BarMagMilTro06a} and the P systems~\cite{Pau02}. Finally, in Section \ref{sec_con} we draw conclusions and we discuss some future developments.
\section{Core Calculus}\label{sec_def}
Term-rewriting formalisms~\cite{BarMagMilTro06a,CCD04,DanLan04,Pau02} have been applied to modelling biological systems. They are characterized by the syntax of terms and the operational semantics. A term represents the structure of the modelled system, and the reduction rules represent the possible evolutions of the system. Some term-rewriting formalisms embed the rules in the terms, other prefer to divide them.\\
In our core calculus, a class contains methods (encapsulation) and extends another class (subtyping); a class inherits all the methods of the class it extends (inheritance). Methods are formed by a sequence of variables (the arguments) and a sequence of reduction rules, expressed in the syntax of the term-rewriting formalism, containing these variables. The methods are called on values of the model, i.e. the biological entities, with a sequence of values as arguments (method invocation). The method invocations are replaced by the reduction rules which are method bodies, where the variables are replaced by the values used as arguments. These reduction rules are then used for the evolution of the model.\\
For the sake of generality, in running examples we use the biological rule notation to represent reduction rules: the syntax is depicted in Figure \ref{fig_SynRule}. We use the notation $\E_1 \tolr{}{} \E_2$ instead of the pair of reduction rules $\E_1 \tol{} \E_2$ and $\E_2 \tol{} \E_1$.\\
\begin{figure}[t]
\[
\begingroup
\begin{array}{lcl}
\E & \eq & \multicolumn{1}{r}{\quad \quad \textit{element composition}}\\
& & \genVariable{v} \  \pipe \  \genVariable{x} \  \pipe \  \E + \E\\
\R & \eq & \multicolumn{1}{r}{\textit{rule declaration}}\\
& & \E \tol{} \E\\
\end{array}
\endgroup
\]
\caption{Biological Rules}
\label{fig_SynRule}
\end{figure}
For example, the hypothetical class of glycoside hydrolase contains a method to hydrolyse a sugar into two sugars, all of them passed as arguments. This method contains the sequence of reduction rules that models hydrolysis. We assign to lactase the glycoside hydrolase type, and then call on it the hydrolysis method, passing as arguments the lactose and the sugar products. By invocation, we obtain the reduction rules specific for lactase, that will be used for the evolution of the model.\\
In this section we present the formal definition of the calculus. The syntax, definitions and rules of the calculus are inspired by the ones proposed by Igarashi, Pierce and Wadler for Featherweight Java ~\cite{IPW01}, a minimal core calculus for modelling the Java Type System. 

\subsection{Syntax}
\begin{figure}[h!]
\[
\begingroup
\begin{array}{lcl}
\multicolumn{3}{c}{Syntax}\\
\CTt & \eq & \multicolumn{1}{r}{\textit{class table declaration}}\\
& & \multiple{\CL}\\
\CL & \eq & \multicolumn{1}{r}{\textit{class declaration}}\\
& & \clss{\genClass{C}}{\genClass{D}}{\multiple{\M}} \quad \quad (\genClass{C}\neq \obj)\\
\M & \eq & \multicolumn{1}{r}{\textit{method declaration}}\\
& & \method{\genMethod{m}}{\multiple{\genClass{C}} \  \multiple{\genVariable{x}}}{\multiple{\R}}\\
\R & \eq & \multicolumn{1}{r}{\textit{reduction rule declaration}}\\
& & \textrm{according to the formalism syntax}\\
& & \textrm{contains variables, values and } \this\\
\I & \eq & \multicolumn{1}{r}{\textit{method invocation}}\\
& & \invoc{\genVariable{v}}{\genMethod{m}}{\multiple{\genVariable{v}}}\\
\multicolumn{2}{l}{\genVariable{x}} & \multicolumn{1}{r}{\textit{variable}}\\
\multicolumn{2}{l}{\genVariable{v}} & \multicolumn{1}{r}{\textit{value}}\\
\multicolumn{2}{l}{\this} & \multicolumn{1}{r}{\textit{this}}\\
\end{array}
\endgroup
\]
\caption{Syntax}
\label{fig_Syn}
\end{figure}
The syntax is given in Figure \ref{fig_Syn}. The metavariables $\genClass{C}$ and $\genClass{D}$ range over class names; $\genMethod{m}$ ranges over method names; $\CL$ ranges over class declarations; $\M$ ranges over method declarations; $\R$ ranges over reduction rules, according to the syntax of the formalism; $\I$ ranges over method invocations; $\genVariable{x}$ ranges over parameter names; $\genVariable{v}$ ranges over values, i.e. the symbols of the model. We assume that the set of variables includes the special variable $\this$. Notice that this is never used as argument of a method.\\
We write $\multiple{\M}$ as shorthand for $\M_1 \ldots \M_n$ and write $\multiple{\genClass{C}}$ for $\genClass{C}_1, \ldots, \genClass{C}_n$ (similarly  $\multiple{\genVariable{x}}$, $\multiple{\genVariable{v}}$, etc.). We abbreviate operations on pairs of sequences similarly, writing $\multiple{\genClass{C}} \  \multiple{\genVariable{x}}$ for $\genClass{C}_1 \  \genVariable{x}_1, \ldots, \genClass{C}_n \  \genVariable{x}_n$, where $n$ is the length of $\multiple{\genClass{C}}$ and $\multiple{\genVariable{x}}$. Sequences of parameter names and method declarations are assumed to contain no duplicate names.\\
The declaration $\clss{\genClass{C}}{\genClass{D}}{\multiple{\M}}$ introduces a class named $\genClass{C}$ with superclass $\genClass{D}$. The new class has the suite of methods $\multiple{\M}$. The methods declared in $\genClass{C}$ are added to the ones declared by $\genClass{D}$ and its superclasses, and may override methods with the same names that are already present in $\genClass{D}$, or add new functionalities. The class $\obj$ has no methods and does not have superclasses.\\
The method declaration $\method{\genMethod{m}}{\multiple{\genClass{C}} \  \multiple{\genVariable{x}}}{\multiple{\R}}$ introduces a method named $\genMethod{m}$ with parameters $\multiple{\genVariable{x}}$ of types $\multiple{\genClass{C}}$. The body of the method is a sequence of reduction rules $\multiple{\R}$, expressed in the syntax of the formalism. The variables $\multiple{\genVariable{x}}$ and the special variable $\this$ are bound in $\multiple{\R}$.\\
A class table $\CTt$ is a mapping from class names $\genClass{C}$ to class declarations $\CL$. We assume a fixed class table $\CTt$ satisfying some sanity conditions: $(1)$ $\CT{\genClass{C}} = \texttt{class } \genClass{C} \ldots$ for every $\genClass{C} \in \dom{\CTt}$; $(2)$ $\obj \notin \dom{\CTt}$; $(3)$ for every class name $\genClass{C}$ (except $\obj$) appearing in $\CTt$, we have $\genClass{C} \in \dom{\CTt}$; $(4)$ there are no cycles in the subtype relation induced by $\CTt$, i.e. a class cannot extends one of its subclasses.\\
The fixed type environment $\G$ contains the association between values $\genVariable{v}$ and their types $\genClass{C}$, written $\typeAss{\genVariable{v}}{\genClass{C}}$. We assume that $\G$ satisfies some sanity conditions: $(1)$ if $\typeAss{\genVariable{v}}{\genClass{C}} \in \G$ for some $\genVariable{v}$, then $\genClass{C} \in \dom{\CTt}$; $(2)$ every value in the set of values (according to the formalism specifications) is associated to exactly one type in $\G$.\\
For example, we define the class of molecules as follows:
\[
\clss{\mol}{\obj}{}
\]
The $\mol$ class has the $\obj$ class as superclass, and it does not have methods, i.e. molecules do not have any particular behaviour.\\
An enzyme is a protein that catalyse chemical reactions. In an enzymatic reaction, the molecules at the beginning of the process (called substrates) are converted into different molecules (called the products), while the enzyme itself is not consumed by the reaction. We define the class of enzymes as follows:
\[
\begingroup
\begin{array}{l}
\clssdef{\enz}{\obj}\\
\{ \\
\quad \methoddef{\act}{\mol \ S, \  \mol \ P}\\
\qquad {S + \this \to \this  + P}\\
\}\\
\end{array}
\endgroup
\]
For the sake of simplicity, in our example an enzyme extends an object rather than a protein, jumping a hierarchy level. According to the enzyme definition, the only method of the $\enz$ class is $\act$, which converts the variable molecule $S$ (the substrate) into the variable molecule $P$ (the product) in presence of the enzyme (the $\this$ variable). In the rest of the paper, the $\mol$ class and its extensions denotes biological object having no particular behaviour, and the $\enz$ class and its extensions denotes biological object having a behaviour.\\
Class tables and environment types are used to create a triple $(\CTt,\G,P)$, where $P$ is a model designed according to the formalism specifications. In $P$ we use method invocations instead of reduction rules. The class table $\CTt$ and the type set $\G$ are fixed, i.e. they are determined during the model creation and cannot vary during model evolution.\\
\subsection{Auxiliary Definitions}
\begin{figure}[h!]
\[
\begingroup
\begin{array}{c}
\textit{Method type lookup}\\ \\
\prooftree
\CT{\genClass{C}} = \clss{\genClass{C}}{\genClass{D}}{\multiple{\M}} \quad \method{\genMethod{m}}{{\multiple{\genClass{C}}} \  \multiple{\genVariable{x}}}{\multiple{\R}} \in \multiple{\M}
\justifies \mtype{\genMethod{m}}{\genClass{C}} = \multiple{\genClass{C}}
\endprooftree
\\ \\
\prooftree
\CT{\genClass{C}} = \clss{\genClass{C}}{\genClass{D}}{\multiple{\M}} \quad \genMethod{m} \textrm{ is not defined in }\multiple{\M}
\justifies \mtype{\genMethod{m}}{\genClass{C}} = \mtype{\genMethod{m}}{\genClass{D}}
\endprooftree
\\ \\
\textit{Method body lookup}
\\ \\
\prooftree
\CT{\genClass{C}} = \clss{\genClass{C}}{\genClass{D}}{\multiple{\M}} \quad \method{\genMethod{m}}{{\multiple{\genClass{C}}} \  \multiple{\genVariable{x}}}{\multiple{\R}} \in \multiple{\M}
\justifies \mbody{\genMethod{m}}{\genClass{C}} = \pair{\multiple{x}}{\multiple{\R}}
\endprooftree
\\ \\
\prooftree
\CT{\genClass{C}} = \clss{\genClass{C}}{\genClass{D}}{\multiple{\M}} \quad \genMethod{m} \textrm{ is not defined in }\multiple{\M}
\justifies \mbody{\genMethod{m}}{\genClass{C}} = \mbody{\genMethod{m}}{\genClass{D}}
\endprooftree
\\
\end{array}
\endgroup
\]
\caption{Auxiliary Definitions}
\label{fig_AuxDef}
\end{figure}
For the typing and evaluation of rules, we need a few auxiliary definitions: these are given in Figure \ref{fig_AuxDef}.\\
The type of a method $\genMethod{m}$ in a class $\genClass{C}$, written $\mtype{\genMethod{m}}{\genClass{C}}$, is a sequence of types $\multiple{\genClass{C}}$. The sequence gives the types of the arguments of the method $\genMethod{m}$ defined in the class $\genClass{C}$, or in one of its superclasses, if not defined in $\genClass{C}$. For example,
\[
\mtype{\act}{\enz} \ = \  \pair{\mol}{\mol}
\]
The body of a method $\genMethod{m}$ in a class $\genClass{C}$, written $\mbody{\genMethod{m}}{\genClass{C}}$, is a pair $\pair{\multiple{\genVariable{x}}}{\multiple{\R}}$ of a sequence of variables $\multiple{\genVariable{x}}$ and a sequence of reduction rules $\multiple{\R}$. The elements of the pair are the arguments and the reduction rules of the method $\genMethod{m}$ defined in the class $\genClass{C}$, or in one of its superclasses, if not defined in $\genClass{C}$. For example,
\[
\mbody{\act}{\enz} \ = \  \pair{\pair{S}{P}}{S + \this \to \this  + P}
\]

\subsection{Evaluation}

\begin{figure}[h!]
\[
\begingroup
\begin{array}{c}
\textit{Method Invocation}\\ \\
\prooftree
\typeAss{\genVariable{v}}{\genClass{C}} \in \G \quad \mbody{\genMethod{m}}{\genClass{C}} = \pair{\multiple{\genVariable{x}}}{\multiple{\R}}
\justifies \invoc{\genVariable{v}}{\genMethod{m}}{\multiple{\genVariable{t}}} \to [\multiple{\genVariable{x}} \mapsto \multiple{\genVariable{t}}, \  \this \mapsto \genVariable{v}] \multiple{\R}
\using \ruleName{e-meth}
\endprooftree
\end{array}
\endgroup
\]
\caption{Evaluation}
\label{fig_Eva}
\end{figure}

The unique evaluation rule concerns the method invocation $\invoc{\genVariable{v}}{\genMethod{m}}{\multiple{\genVariable{t}}}$. In this case, if the value $\genVariable{v}$ has type $\genClass{C}$ in $\G$, and the method $\genMethod{m}$ has arguments $\multiple{\genVariable{x}}$ and body $\multiple{\R}$ in $\genClass{C}$, then its evaluation is the sequence of reduction rules $\multiple{\R}$, in which all the occurrences of the variables $\multiple{\genVariable{x}}$ are replaced with the values $\multiple{\genVariable{t}}$, and all the occurrences of $\this$ are replaced with the value $\genVariable{v}$. A method invocation is placed in the model instead of the reduction rules: once evaluated, the reduction rules of the method become the reduction rules of the model.\\
Phosphoglucose isomerase is an enzyme that catalyses the conversion of glucose-6-phosphate into fructose 6-phosphate (and vice versa) in the second step of glycolysis. In order to model this behaviour, in $\G$ we associate to the value $\phoiso$ (the phosphoglucose isomerase) the type $\enz$, and to the values $\glu$ and $\fru$ (the glucose-6-phosphate and fructose 6-phosphate, respectively) the type $\mol$
\[
\G = \{\typeAss{\phoiso}{\enz}, \typeAss{\glu}{\mol}, \typeAss{\fru}{\mol}\}
\]
Instead of the reduction rules, in the model we place the calling of the $\act$ method on the $\phoiso$ enzyme, using the molecules as arguments
\[
\invoc{\phoiso}{\act}{\glu,\fru}
\]
Following the evaluation rule in Figure \ref{fig_Eva}, this method invocation is replaced by the reduction rule
\[
\glu + \phoiso \to \phoiso  + \fru
\]
As a consequence, we obtain the reduction rule modelling the conversion of glucose-6-phosphate into fructose 6-phosphate. In order to obtain the conversion in the other side, we call the $\act$ method on the $\phoiso$ enzyme swapping the arguments
\[
\invoc{\phoiso}{\act}{\fru,\glu}
\]
This method invocation is then replaced by the reduction rule
\[
\fru + \phoiso \to \phoiso  + \glu
\]
After method evaluation, we obtain the reduction rules of the model, representing the possible evolution of the system.
\subsection{Typing}

\begin{figure}[h!]
\[
\begingroup
\begin{array}{c}
\textit{Subtyping}\\ \\
\subtype{\genClass{C}}{\genClass{C}} \quad \ruleName{t-sub1} \quad \quad
\prooftree
\subtype{\genClass{C}}{\genClass{D}} \quad \subtype{\genClass{D}}{\genClass{E}}
\justifies \subtype{\genClass{C}}{\genClass{E}}
\using \ruleName{t-sub2}
\endprooftree
\\ \\
\prooftree
\CT{\genClass{C}} = \clss{\genClass{C}}{\genClass{D}}{\multiple{\M}}
\justifies \subtype{\genClass{C}}{\genClass{D}}
\using \ruleName{t-sub3}
\endprooftree
\end{array}
\endgroup
\]
\caption{Subtyping}
\label{fig_Sub}
\end{figure}

The rules for subtyping are formally defined in Figure \ref{fig_Sub}. The subtype relation between classes is given by the class declarations in the class table $\CTt$. The subtype relation is reflexive and transitive. For $\enz$ and $\mol$ classes we derive the following subtype relations:
\[
\begingroup
\begin{array}{ccc}
\subtype{\enz}{\enz} & \subtype{\mol}{\mol} & \textrm{(by rule } \ruleName{t-sub1} \textrm{)}\\
\subtype{\enz}{\obj} & \subtype{\mol}{\obj} & \textrm{(by rule } \ruleName{t-sub3} \textrm{)}
\end{array}
\endgroup
\]
Note that $\enz$ is not a subtype of $\mol$, then an enzyme cannot be a substrate nor a product of the $\enz$'s $\act$ method.\\
\begin{figure}[h!]
\[
\begingroup
\begin{array}{c}
\textit{Invocation typing}\\ \\
\prooftree
\typeAss{\genVariable{v}}{\genClass{C}} \in \G \quad \mtype{\genMethod{m}}{\genClass{C}} = \multiple{\genClass{C}} \quad \typeAss{\multiple{\genVariable{t}}}{\multiple{D}}\in \G \quad \subtype{\multiple{D}}{\multiple{C}}
\justifies \invoc{\genVariable{v}}{\genMethod{m}}{\multiple{\genVariable{t}}} \okIn{\genClass{C}}
\using \ruleName{t-invmeth}
\endprooftree
\\ \\
\textit{Method typing}\\ \\
\prooftree
\typeAss{\multiple{\genVariable{x}}}{\multiple{\genClass{C}}}, \typeAss{\this}{\genClass{C}} \vdash \multiple{\R} \ok
\justifies  \method{\genMethod{m}}{{\multiple{\genClass{C}}} \  \multiple{\genVariable{x}}}{\multiple{\R}} \okIn{\genClass{C}}
\using \ruleName{t-clmeth}
\endprooftree
\\ \\
\textit{Class typing}\\ \\
\prooftree
\CT{\genClass{C}} = \clss{\genClass{C}}{\genClass{D}}{\multiple{\M}} \quad \multiple{\M} \okIn{\genClass{C}}
\justifies \clss{\genClass{C}}{\genClass{D}}{\multiple{\M}} \ok
\using \ruleName{t-class}
\endprooftree
\\
\end{array}
\endgroup
\]
\caption{Typing}
\label{fig_Typ}
\end{figure}

The typing rules for method invocations and for method and class declarations are given in Figure \ref{fig_Typ}. Typing statements for method invocations have the form $\invoc{\genVariable{v}}{\genMethod{m}}{\multiple{\genVariable{t}}} \ok$, asserting that the method invocation $\invoc{\genVariable{v}}{\genMethod{m}}{\multiple{\genVariable{t}}}$ is well formed. The typing rule checks that the types of the values used as arguments in a method invocation are subtypes of the types of the arguments required by the method.\\
Typing statements for method declarations have the form $\M \okIn{\genClass{C}}$, and assert that the method declaration $\M$ is well formed in the class $\genClass{C}$. The typing rule checks that the reduction rules in the method of a class are well formed, according to the types of the arguments and the class. The relation $\vdash$ serves this purpose, by using the type assignments on its left in the type checking of the element on its right. Different formalisms have different constraints to check if a rule is well formed. For this reason, the modeller must add the proper typing rules to check the well-formedness of a rule, according to the types of the arguments and the class in which it is contained, in addition to the typing rules in Figure \ref{fig_Typ}.\\
Typing statements for class declarations have the form $\CL \ok$, stating that the class declaration $\CL$ is well formed. The typing rule checks that each method declaration in the class is well formed.\\
As expected, both $\enz$ and $\mol$ classes are $\ok$.\\
Note that the inheritance is very simple: a class inherits all the methods of its superclass, and it can modify the body and the arguments of a method declared in its superclass, i.e. it can change the reduction rules and the arguments associated to a method name. In this way, lower classes can reuse the names of higher classes methods, i.e. more specialised biological entities can focus and specialise the behaviour of more generic biological entities by reusing the name associated to a generic reduction rule. For example, an hydrolase is an enzyme, but it cannot catalyse any reaction except hydrolysis. For this reason, we design hydrolase class as follows:
\[
\begingroup
\begin{array}{l}
\clssdef{\hyd}{\enz}\\
\{ \\
\quad \methoddef{\act}{\mol \ S, \  \mol \ P_1, \  \mol \ P_2}\\
\qquad S + H_2O + \this \to \this  + P_1 + P_2\\
\}\\
\end{array}
\endgroup
\]
The $\hyd$ class is an extension of the $\enz$ class that overrides the $\act$ method. In this way, the generic catalysis described in the $\enz$'s $\act$ method is no more available in the $\hyd$ class, but the override $\act$ method makes available the specific hydrolysis.\\
In the same way, the glycoside hydrolase is an hydrolase, but its substrate and products are sugars. Then the glycoside hydrolase class is designed as an extension of $\hyd$ class, that overrides the $\act$ method by modifying the types of the arguments, from molecules to sugars:
\[
\begingroup
\begin{array}{l}
\clss{\sug}{\mol}{}\\
\\
\clssdef{\glhyd}{\hyd}\\
\{ \\
\quad \methoddef{\act}{\sug \ S, \  \sug \ P_1, \  \sug \ P_2}\\
\qquad  S + H_2O + \this \to \this  + P_1 + P_2 \\
\}\\
\end{array}
\endgroup
\]

\section{Example}\label{sec_exa}
In this section we show how our calculus can be used to model biological behaviours. As an example, we design classes and method invocations to describe Michaelis-Menten enzyme kinetic, the two-substrates enzyme kinetic and the competitive inhibition kinetic.
\subsection{Michaelis-Menten Model}\label{mmm}
In the Michaelis-Menten Model, the enzyme reaction is divided in two stages. In the first stage, the substrate $S$ binds reversibly to the enzyme $E$, forming the enzyme-substrate complex $ES$, then in the second one the enzyme catalyses the chemical step in the reaction and releases the product $P$:
\[
E + S \tolr{}{} ES \to E + P
\]
This basic behaviour is also used in most complex enzyme reactions. In order to model this behaviour, we create two classes, the $\enz$ class and the $\enzcomp$ class. The first one models an enzyme: it associates itself with a substrate and produces an enzyme-substrate complex. The second one models an enzyme-substrate complex: it dissociates itself in an enzyme and a product.
\[
\begingroup
\begin{array}{ll}
\clssdef{\enz}{\obj} & \clssdef{\enzcomp}{\enz}\\
\{ & \{ \\
\quad \methoddef{\ass}{\mol \ S, \  \enzcomp \ ES} & \quad \methoddef{\dis}{\enz \ E, \  \mol \ P}\\
\qquad S + \this \to ES & \qquad \this \to E  + P \\
\} & \}
\end{array}
\endgroup
\]
Since an enzyme-substrate complex can act as an enzyme, the $\enzcomp$ class extends the $\enz$ class. In this way, the $\enzcomp$ class inherits from $\enz$ the $\ass$ method by auxiliary definitions.\\
The type environment is
\[
\G = \{\typeAss{E}{\enz}, \typeAss{ES}{\enzcomp}, \typeAss{S}{\mol}, \typeAss{P}{\mol}\}
\]
The method invocations for reproducing the described behaviour are
\[
\invoc{E}{\ass}{S,ES} \quad \quad \invoc{ES}{\dis}{E,S} \quad \quad \invoc{ES}{\dis}{E,P}
\]

\subsection{Two-substrates Enzymes}
Some enzymes catalyse reaction between two substrates. This reaction is usually divided into three stages. In the first, the substrate $S_1$ binds reversibly to the enzyme $E$, forming the enzyme-substrate complex $ES_1$, then in the second the substrate $S_2$ binds reversibly to the enzyme-substrate complex $ES_1$, forming the enzyme-substrate complex $ES_1S_2$. Finally the enzyme complex $ES_1S_2$ catalyses the chemical step in the reaction and releases the product $P$:
\[
E + S_1 \tolr{}{} ES_1 \  + S_2 \tolr{}{} ES_1S_2 \to E + P
\]
Note that this is only one of all the possible interactions between an enzyme and two substrates. To model this behaviour, we assign the following types:
\[
\begingroup
\begin{array}{cl}
\G = & \{\typeAss{E}{\enz}, \typeAss{ES_1}{\enzcomp}, \typeAss{ES_1S_2}{\enzcomp}, \\
& \typeAss{S_1}{\mol}, \typeAss{S_2}{\mol}, \typeAss{P}{\mol}\}\\
\end{array}
\endgroup
\]
The method invocations are the following:
\[
\begingroup
\begin{array}{c}
\invoc{E}{\ass}{S_1,ES_1} \quad \quad \invoc{ES_1}{\dis}{E,S} \quad \quad \invoc{ES_1}{\ass}{S_2,ES_1S_2}  \\
\invoc{ES_1S_2}{\dis}{ES_1,S_2} \quad \quad \invoc{ES_1S_2}{\dis}{E,P}
\end{array}
\endgroup
\]
\subsection{Competitive Inhibition}
In Biology, enzyme reaction rates can be decreased by molecules called enzyme inhibitors. There exist a lot of inhibitors kinetics: among others, in Competitive Inhibition the inhibitor $I$ binds to enzyme $E$ producing the complex $EI$ and stops a substrate $S$ from entering the enzyme's active site and producing the complex $ES$. The inhibitor and substrate compete for the enzyme (i.e. they cannot bind at the same time):
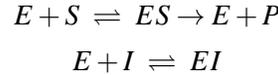
\[
\begingroup
\begin{array}{c}
E + S \tolr{}{} ES \to E + P\\
E + I \tolr{}{} EI
\end{array}
\endgroup
\]
This case is an extension of the Michaelis-Menten Model in Section \ref{mmm}, and is modelled by adding the following type environment and method invocations:
\[
\begingroup
\begin{array}{c}
\G' = \{\typeAss{EI}{\enzcomp}, \typeAss{I}{\mol}\}\\
\invoc{E}{\ass}{I,EI} \quad \quad \invoc{EI}{\dis}{E,I}
\end{array}
\endgroup
\]

\section{Use of Classes in Term-Rewriting Formalisms}\label{sec_use}
The calculus in this paper aims to be easily applicable to the most popular term-rewriting formalisms for modelling biological systems. To do so, we just act as follows:
\begin{enumerate*}
 \item set the syntax of reduction rules of the term-rewriting formalism as the syntax of reduction rules of the core calculus;
 \item if the reduction rules must respect certain conditions handled by typing, then add the proper typing rules to check their well-formedness;
 \item define the class table $\CTt$ and assign types to values in the type environment $\G$ according to their biological behaviour;
 \item create a triple $(\CTt,\G,P)$, where $P$ is a model designed according to the formalism specifications, except for the reduction rules, that are replaced by method invocations.
\end{enumerate*}
After the evaluation of the method invocations in $P$, we obtain the model $P'$ in the formalism form, in which all the reduction rules are consistent with the biological classification and behaviour defined in $\CTt$ and $\G$.\\

We present an implementation of the calculus in two different term-rewriting formalisms: the Calculus of Looping Sequences (CLS) and the P systems. As case study, we present the Porins behaviour. Porins are proteins that cross a cellular membrane and act as a pore through which molecules can diffuse. The molecules which diffuse across the porin depends on the porin itself. Among the porins, aquaporins selectively conduct water molecules in and out of the cell, while preventing the passage of ions and other solutes. Some of them, known as aquaglyceroporins, transport also other small uncharged solutes, such as glycerol, CO2, ammonia and urea across the membrane (see~\cite{HD08}). We design the $\por$ class to model the porin behaviour, and we present an example of triple and its evaluation, in CLS and P systems formalisms. In particular, we model two kinds of aquaporins: one kind transports only water, the other one transports both urea and water.
\subsection{Calculus of Looping Sequences}
A CLS model~\cite{BarMagMilTro06a} is composed by:
\begin{itemize*}
 \item a set $\EE$ of elements;
 \item sets $\XX$, $\SV$ and $\TV$ of element, sequence and term variables, respectively;
 \item a set $\RR$ of reduction rules (called \textit{rewrite rules}) in the form $P \to P$, according to the pattern syntax in Figure \ref{fig_SynCLS};
 \item a term $T$, i.e. a pattern without variables.
\end{itemize*}

\begin{figure}[h!]
\[
\begingroup
\begin{array}{lcl}
P\; & ::= SP \agr \Loop{SP} \into P \agr P \pipe P \agr X\\
SP\; & ::= \epsilon \agr a  \agr SP \cdot SP \agr \xx \agr x\\
C\; & ::= \phole \agr C \pipe T \agr T \pipe C \agr \Loop{S} \into C
\end{array}
\endgroup
\]
\caption{Syntax of Patterns, Sequence Patterns and Contexts in CLS}
\label{fig_SynCLS}
\end{figure}
A rewrite rule $P_1 \to P_2$ states that a term $P_1\sigma$, obtained by instantiating variables in $P_1$ by some instantiation function $\sigma$, a function that maps variables to terms preserving the kind of the variables, can be transformed into the term $P_2\sigma$. According to the context syntax in Figure \ref{fig_SynCLS}, the term $C[P_1\sigma]$ evolve in the term $C[P_2\sigma]$ by rewrite rule $P_1 \to P_2$, where $C[T]$ denotes the term obtained by replacing the unique $\phole$ with $T$ in $C$.\\
Since in $CLS$ the reduction rules have the form $P \to P$, the rule syntax of the classes becomes
\[
\R \eq P \to P.
\]
A model is a pair $\pair{T}{\RR}$, where $T$ is the term depicting the initial state of the system, and $\RR$ is the set of rewrite rules. Using classes and methods, the set $\RR$ becomes a set of method invocations, $\RR = \{\multiple{I}\}$, which must be evaluated in an initial phase of system initialisation, before the evaluation of the term, to obtain the rewrite rules of the model.\\ 
A class modelling the porin behaviour with rewrite rules in CLS syntax is the following:
\[
\begingroup
\begin{array}{l}\clssdef{\por}{\obj}\\
\{ \\
\quad \methoddef{\inm}{\mol \ S}\\
\qquad S \pipe \Loop{\this \cdot \xx}\into X \to \Loop{\this \cdot \xx}\into (S \pipe X) \\\\
\quad \methoddef{\outm}{\mol \ S}\\
\qquad \Loop{\this \cdot \xx}\into (S \pipe X) \to S \pipe \Loop{\this \cdot \xx}\into X \\
\}\\
\end{array}
\endgroup
\]
We use the symbols $w$ for water, $u$ for urea, $AW$ for the aquaporin that transports only water and $AWU$ for the aquaporins that transports both water and urea. In our term, both kinds of aquaporins are included into a membrane:
\[
T = w \pipe \ldots \pipe w \pipe u \pipe \ldots \pipe u \pipe \Loop{AW}\into(\epsilon) \pipe \Loop{AWU}\into(\epsilon)
\] 
The type environment is the following:
\[
\begingroup
\begin{array}{cl}
\G = & \{\typeAss{AW}{\por}, \typeAss{AWU}{\por}, \typeAss{w}{\mol}, \typeAss{u}{\mol}\}\\
\end{array}
\endgroup
\]
and the class table $\CTt$ contains the $\por$ and $\mol$ classes. The triple is $(\CTt,\G,P)$, where $P$ is composed by the term $T$ and the rule set containing the following method invocations:
\[
\begingroup
\begin{array}{ccc}
\invoc{AW}{\inm}{w} &\quad \invoc{AW}{\outm}{w} & \quad \invoc{AWU}{\inm}{w}\\
\invoc{AWU}{\outm}{w} &\quad \invoc{AWU}{\inm}{u} & \quad \invoc{AWU}{\outm}{u}\\
\end{array}
\endgroup
\]
After the evaluation of the triple, the CLS model is composed by the term $T$ and the rewrite rules
\[
\begingroup
\begin{array}{cc}
w \pipe \Loop{AW \cdot \xx}\into X \to \Loop{AW \cdot \xx}\into (w \pipe X) \quad & \quad \Loop{AW \cdot \xx}\into (w \pipe X) \to w \pipe \Loop{AW \cdot \xx}\into X\\
w \pipe \Loop{AWU \cdot \xx}\into X \to \Loop{AWU \cdot \xx}\into (w \pipe X) \quad & \quad \Loop{AWU \cdot \xx}\into (w \pipe X) \to w \pipe \Loop{AWU \cdot \xx}\into X\\
u \pipe \Loop{AWU \cdot \xx}\into X \to \Loop{AWU \cdot \xx}\into (u \pipe X) \quad & \quad \Loop{AWU \cdot \xx}\into (u \pipe X) \to u \pipe \Loop{AWU \cdot \xx}\into X\\
\end{array}
\endgroup
\]

\subsection{P systems}
A P-system~\cite{Pau02} is a n-tuple $\Pi = (V, \mu, M_1, \ldots, M_n, (R_1,\rho_1), \ldots, (R_n,\rho_n), i_0)$, where
\begin{itemize*}
 \item $V$: alphabet;
 \item $\mu$: membrane structure of degree $n$, with the membrane and the regions labelled in a one-to-one manner with elements in a given set $\labelSet$;
 \item $M_i$: multisets of symbols (or strings) in $V$, the symbols contained in the membrane $i$;
 \item $R_i$: finite sets of reduction rules (called \textit{evolution rules}) $x \to y$ contained in the membrane $i$ and such that $x \in V^{*}$ and $y = y'$ or $y = y'\delta$, where $y' \in (V \times \{here, out\})^{*} \cup (V \times \{in_j \pipe j \in \labelSet \})^{*}$;
 \item $\rho_i$: partial order relations over $R_i$;
 \item $i_0$: a label in $\labelSet$ which specifies the output membrane. If empty, then the output region is the environment.
\end{itemize*}
Consider an evolution rule $x \to y$ in the set $R_i$: if the symbols in $x$ appear in $M_i$, then these symbols are replaced by the symbols in $y$ according to the rule. If a symbol $a$ appears in $y$ in a pair $(a, here)$, then it will remain in $M_i$. If a symbol $a$ appears in $y$ in a pair $(a, out)$, then it becomes a symbol of the membrane immediately outside the membrane $i$, according to the membrane structure $\mu$. If a symbol $a$ appears in $y$ in a pair $(a, in_j)$, and the membrane $j$ is contained in the membrane $i$ according to the membrane structure $\mu$, then it becomes a symbol of the membrane $j$. If $y = y'\delta$, then the membrane $i$ and the evolution rules in $R_i$ disappear, and all the symbols in $M_i$ are added to the symbols of the membrane immediately outside the membrane $i$. Evolution rules are applied following the priority in $\rho_i$, and in a non-deterministic way in case of same priority. In a single evolution step, all symbols in all membranes evolve in parallel, and every applicable evolution rule is applied as many times as possible.\\
According to the definitions of evolution rules, the rule syntax becomes
\[
\R \eq x \to y
\]
Using classes and methods, each set $R_i$ becomes a set of method invocations, $R_i=\multiple{I_i}$.\\
In P systems we have two kinds of symbols which may be involved in an evolution rule: the biological entities (contained in $V$) and the labels of membranes (contained in $\labelSet$). Since they are different entities, we must design a distinct class for everyone of them. As a solution, we construct the class $\bioobj$ for biological entities, and $\lab$ for labels, both extending $\obj$.
\[
\begingroup
\begin{array}{l}\clss{\bioobj}{\obj}{}\\
\clss{\lab}{\obj}{}\\
\end{array}
\endgroup
\]
All the biological entities must extend $\bioobj$ or one of its subclasses. For example, the definition of the class $\mol$ is
\[
\clss{\mol}{\bioobj}{}
\]
A class modelling the porin behaviour with P-system evolution rules is the following:
\[
\begingroup
\begin{array}{l}\clssdef{\por}{\bioobj}\\
\{ \\
\quad \methoddef{\inm}{\mol \ S, \lab \ J}\\
\qquad S  \to S (in_J) \\\\
\quad \methoddef{\outm}{\mol \ S}\\
\qquad S  \to S (out) \\
\}
\end{array}
\endgroup
\]
In this case, the aquaporin that transports only water ($w$) is contained into the membrane labelled by $1$, and the other one, that transports both urea ($u$) and water, is contained into the membrane labelled by $2$. The type environment is the following:
\[
\begingroup
\begin{array}{cl}
\G = & \{\typeAss{A}{\por}, \typeAss{w}{\mol}, \typeAss{u}{\mol}, \typeAss{0}{\lab}, \typeAss{1}{\lab}, \typeAss{2}{\lab}\}\\
\end{array}
\endgroup
\]
and the class table $\CTt$ contains the $\por$, $\mol$ and $\bioobj$ classes. The triple is $(\CTt,\G,\Pi)$, where $\Pi$ is the following:
\[
\begingroup
\begin{array}{cl}
\Pi = &(\{u,w,A\}, [[]_2[]_3]_1, \{u,\ldots u,w,\ldots,w\}, \emptyset, \emptyset, (\invoc{A}{\inm}{w,1},\invoc{A}{\inm}{w,2},\\
& \invoc{A}{\inm}{u,2}),(\invoc{A}{\outm}{w}), (\invoc{A}{\outm}{w},  \invoc{A}{\outm}{u}), 1)
\end{array}
\endgroup
\]
After the evaluation of the method invocations, we obtain the P-system
\[
\begingroup
\begin{array}{cl}
\Pi' = &(\{u,w,A\}, [[]_2[]_3]_1, \{u,\ldots u,w,\ldots,w\}, \emptyset, \emptyset, (w  \to w (in_1), w \to w (in_2),\\
&  u \to u (in_2)),(w \to w (out)), (w \to w (out), u \to u (out)), 1)
\end{array}
\endgroup
\]

\section{Conclusions and Future Developments}\label{sec_con}
Modularity is the key idea to manage the complexity of biological processes, because it allows molecules or compartments to be specified and then combined. It is usually combined with abstraction, that allows generic properties to be specified independently of specific instances: the result are parametrised modules. These are widely used in formalisms designed to model biological systems: for example, P-Lingua~\cite{GGPPR08} is a programming language for membrane computing which aims to be a standard to define P systems. A P-Lingua program consists of a set of parametrised programming modules composed by a sequence of sentences in P-lingua: these sentences are the membrane structure of the model or the rules and objects contained into these membranes. Modules are executed by using calls, that assign some values to their parameters.\\
Modules, in particular if parametrised, permit to define a structure and re-use it, but they have a limitation: they are applicable to every molecule, without limitations, while usually modules are designed only for some kinds of molecules. To manage this problem, some formalisms add a simple Type System to modularity and abstraction: this Type System just checks the correspondence between the types of the arguments and the types of the parameters in a module call operation. Biochemical Systems (LBS)~\cite{PP10} combine rule-based approaches to modelling with modularity. Modules may be parametrised on compartments, rates, and species. Species are typed by the names of their component atomic species and of their modification site types: when a method is called, the Type System checks the correspondence between the types of the arguments and the types of the parameters. A simple Type System is also implemented in Little b~\cite{littleb}, a high-level programming language for modular model building. In Little b a modeller can define monomers, composed by a name and a sequence of bond sites: these can connect each other by labelling their bond sites, creating complexes; reactions are pairs of patterns that specify the transformation of complexes matching the first pattern to the second one, and may create or delete links between sites. Sites can be labelled with tags, that specify the kind of link of the site and the kind of links it accepts: this tag-based system serves as Type System, and in particular as a type checker.\\
All the above samples do not let to specify a hierarchy between the typed objects (species for LBS and sites for little b): a hierarchic structure permits more advanced tools and analyses. An example of use of hierarchy to manage the complexity of biological system is the extension of Kappa with agent hierarchies~\cite{DFHK09}. A Kappa model consists of a collection of rules and agents; each agent has an associated set of sites. Modellers can define variants on an agent by adding or replacing its sites: the variance relation create an agent hierarchy. A generic rule is then expanded into a set of concrete rules by replacing each agent in the rule with all appropriate agents below it in the hierarchy: so the hierarchy is used with the purpose to enable rapid development of large rule sets via the mechanism of generic rules. Moreover, the same hierarchic structure is used for a static analysis of the rule set: an analyser navigates the space of variants of a model looking if, with the current rule set, a specific concrete rule can or cannot take place under a sequence of conditions. Even if this procedure can never prove that a rule is correct, it can be used to reject rules that lead to behaviour incompatible with experimental results.\\
Our calculus takes advantage of modularity, abstraction and hierarchy by constructing a parametrised module hierarchic structure for expressing reduction rules. Using classes instead of modules, our calculus can express the hierarchic structure of Biological ontologies, and also exploit the features of Object-Oriented programming, such as inheritance and subtyping. On the other side, the rules in a class are not visible from outside, then the resolution of the errors becomes more difficult. Finally, our calculus does not specify a meta-language, because it aims to be used with different term-rewriting formalisms: this lack of structure is the more evident difference with the other approaches, but it pays off in terms of expressiveness, because we cannot exploit the expressive power of a particular syntax.\\

Summarizing, modularity allows behaviours to be specified and then combined; hierarchy allows typechecking and re-use of the behaviour; abstraction allows generic properties to be specified independently of specific instances. The modularity, hierarchy and abstraction of the classes enables libraries to be created for generic biological processes, which can be instantiated and re-used repeatedly in different contexts with different arguments. These libraries could be designed and refined by experts, and then made available to all modellers, thereby creating a scientific "commons" for model building. Moreover, they can be used in different models, ensuring that their reduction rules are consistent with the biological ontology defined in them. These libraries could also be adapted from a formalism to another, rewriting the reduction rules and with small alteration to the hierarchy, if needed. That modularity allows the Bioinformatics field to evolve in a decentralized manner, because any user can develop novel abstractions of the biology being studied in any formalism and contribute these back to the community, that can adapt these classes to any particular formalism.\\

The calculus proposed in this paper implements only very basic features of object-oriented paradigm. In the opinion of the author, these features are the most common and useful in biological modelling, but increasing the complexity of the modelled systems the need of new features could emerge. For example, sometimes molecules may have different roles depending on the context: our calculus cannot deal with this behaviour, because every value is associated to exactly one type. For this reason, a possible development is surely the study and implementation of other basic and high-level constructs of imperative and object-oriented paradigms, such as data structures, multiple inheritance or parametric polymorphism (also known as generics).\\

In our calculus, the modeller decide which reduction rules to include in a model, but in this way a raw modeller could forget some important rule. A possible evolution is to infer the reduction rules directly from the composition of the model, according to the association between classes and values defined in the type environment. For example, if the term of the model contains a porin, then the system may infer the proper reduction rules to include, in this case the ones modelling the passage of elements through membranes. Moreover, in this way the reduction rules in a model could become dynamic: they could evolve following the evolution of the model, in a correct (from a biological point of view) way, without any external intervention. For example, if, during the evolution of the model, a lactase is created in the term, then the type system may add the proper reduction rules, in this case the ones modelling hydrolysis.

\paragraph{Acknowledgements.} The author thanks the referees for their helpful comments.
The final version of the paper improved due to their suggestions, in particular the Section \ref{sec_con}.

\bibliographystyle{eptcs}

\end{document}